\documentclass[authoryear,preprint,12pt]{elsarticle}

\usepackage[T1]{fontenc}
\usepackage{lmodern}
\usepackage{amssymb,amsmath}
\usepackage{graphicx}
\usepackage{booktabs}
\usepackage{tabularx}
\usepackage{array}
\usepackage{placeins}
\usepackage{microtype}
\usepackage[colorlinks=true,allcolors=blue,bookmarks=false]{hyperref}
\usepackage{orcidlink}
\usepackage{lineno}
\usepackage{geometry}
\newcommand{\bperp}{\beta_{\parallel p}}
\newcommand{\Rp}{R_p}
\newcommand{\Op}{\Omega_p}

\journal{Journal of Subatomic Particles and Cosmology}
\myfooter[L]{Accepted for publication in Journal of Subatomic Particles and Cosmology}

\graphicspath{{./}{figures/}}

\newcommand{\safeincludegraphics}[2][]{%
  \IfFileExists{#2}{\includegraphics[#1]{#2}}{%
    \fbox{%
      \begin{minipage}[c][0.25\textheight][c]{0.88\linewidth}
      \centering
      Missing figure file: \texttt{\detokenize{#2}}\\[0.5em]
      \end{minipage}%
    }%
  }%
}

\begin{document}

\begin{frontmatter}

\title{Alfv\'enicity and Proximity to Parallel-Mode Marginal Stability
       in the Slow Solar Wind}

\author[1]{Mani~K~Chettri\fnref{fn1}\texorpdfstring{\,\orcidlink{0009-0000-1368-9263}}{}}

\author[1]{Rupak~Mukherjee\texorpdfstring{\,\orcidlink{0000-0003-3955-7116}}{}}

\author[2]{Hemam~D.~Singh\fnref{fn2}\texorpdfstring{\,\orcidlink{0009-0003-3061-8944}}{}}
\fntext[fn1]{Email: mkchettri8@gmail.com}
\fntext[fn2]{Corresponding author: hemam.singh@nsut.ac.in}

\affiliation[1]{organization={Department of Physics, Sikkim University},
            city={Gangtok},
            postcode={737102},
            country={India}}
\affiliation[2]{organization={Department of Physics, Netaji Subhas University
            of Technology},
            addressline={Sector 3, Dwarka},
            city={New Delhi},
            postcode={110078},
            country={India}}

\begin{abstract}
The proton temperature anisotropy in the solar wind is bounded by the
thresholds of pressure-anisotropy-driven kinetic instabilities, and the
distance at which the plasma settles from these thresholds is thought to be
regulated by compressive fluctuations through the fluctuating anisotropy
effect. We test whether the level of Alfv\'enicity is associated with this
distance in the slow solar wind. Using five years of Wind/SWE bi-Maxwellian
proton measurements (2004--2008), we separate slow-wind intervals into
Alfv\'enic and non-Alfv\'enic populations on the basis of normalized cross
helicity and residual energy. We compare their magnetic compressive fraction
and normalized compressive amplitude, and measure their proximity to marginal
stability in the plane of parallel proton beta and temperature anisotropy using
the maximum growth rate over the scanned parallel wavenumbers,
$\gamma_{\parallel,\max}$, from a Vlasov dispersion solver. At low matched
parallel beta, the Alfv\'enic slow wind has lower normalized field-strength fluctuation
amplitude and reaches the parallel-mode marginal-stability criterion more often
than the non-Alfv\'enic slow wind. This association is consistent with a weaker
fluctuating-anisotropy effect in the Alfv\'enic slow wind, but does not establish
a causal relation.
\end{abstract}

\begin{keyword}
solar wind \sep turbulence \sep plasma instabilities \sep
kinetic plasmas \sep data analysis
\end{keyword}

\end{frontmatter}

\section{Introduction}
\label{sect:intro}

The solar wind is a weakly collisional, turbulent plasma whose fluctuations
carry a strong imprint of Alfv\'enic correlations between the velocity and the
magnetic field, first recognized in the earliest in situ measurements
\citep{Coleman1968,BelcherDavis1971}. The turbulent cascade, its anisotropy,
and its evolution with heliocentric distance have since been characterized in
detail \citep{TuMarsch1995,BrunoCarbone2013}. Because the plasma is only weakly
collisional, the cascade does not terminate in a simple fluid dissipation
range; instead the fluctuations reach the proton kinetic scales, where they
couple to the particle velocity distributions and help to set the thermodynamic
state of the protons \citep{Marsch2006,Verscharen2019}. Because it can be sampled
in situ, the solar wind is the most accessible collisionless plasma in
which these kinetic processes can be studied directly: turbulent dissipation,
wave--particle interactions, and the regulation of velocity-space anisotropy by
microinstabilities. The same processes operate in other weakly collisional
astrophysical plasmas, such as stellar coronae and the intracluster medium.
Whether this kinetic regulation differs between the distinct types of solar
wind is not well established, and is the question we take up here.

A clear manifestation of kinetic regulation is the bounding of the proton
temperature anisotropy $\Rp=T_\perp/T_\parallel$. Linear kinetic theory predicts
that, at a given parallel proton plasma beta $\bperp$---the ratio of parallel
proton thermal pressure to magnetic pressure---departures from isotropy are
limited by pressure-anisotropy-driven instability thresholds, namely
the proton cyclotron and mirror instabilities for $T_{\perp} > T_{\parallel}$
and the parallel and oblique firehose instabilities for $T_{\parallel} >
T_{\perp}$ \citep{Gary1993,Hellinger2006}. In situ measurements at 1~au show
that the observed anisotropy is largely confined within these thresholds
\citep{Kasper2002,Hellinger2006,Maruca2011,Bale2009}, a constraint that is
already imposed close to the Sun and is maintained as the wind expands
\citep{Matteini2007}, and that the gyroscale magnetic fluctuation power is
enhanced along the marginal stability boundaries \citep{Bale2009}. The relevant
bounding curve depends on the sense of the anisotropy and on beta; on the
$T_{\parallel} > T_{\perp}$ side the oblique firehose provides a tighter bound
than its parallel counterpart only in the high-beta regime
\citep{HellingerTravnicek2008}.

The mean distance at which the plasma settles from these boundaries is not set
by the instabilities alone. \citet{Verscharen2016} showed that large-scale
compressive fluctuations make $\bperp$ and $\Rp$ fluctuate about their mean
values, so that excursions repeatedly cross a threshold, excite the
corresponding instability, and drive partial isotropization. This fluctuating
anisotropy effect holds the mean state a finite distance from marginal
stability, the distance increasing with the amplitude of the compressive
driving. It predicts, in particular, that plasma threaded by weaker compressive
fluctuations should be able to relax closer to the instability thresholds.

The slow solar wind is not a single population \citep{DAmicisBruno2015}. A subset
is highly Alfv\'enic, with strong velocity--magnetic-field correlation and a low
magnetic compressive fraction, and shares many properties with the fast wind
despite its low speed \citep{DAmicis2019,Stansby2020,Perrone2020}. The degree of
Alfv\'enicity is conveniently quantified by the normalized cross helicity and
residual energy \citep{BelcherDavis1971,Bavassano1998}. These quantities
distinguish imbalanced Alfv\'enic streams from more balanced streams, but the
compressive fraction alone does not specify the normalized amplitude of the
field-strength fluctuations.

We therefore compare the normalized compressive amplitudes of the Alfv\'enic and
non-Alfv\'enic slow wind and test how often they reach the parallel-mode
marginal-stability criterion at matched $\bperp$. An association between these
quantities would be consistent with the fluctuating-anisotropy interpretation,
but would not establish causation.

\section{Data and Methods}
\label{sect:data}

\subsection{Dataset and Interval}
\label{sect:dataset}

We use bi-Maxwellian proton fits from the Solar Wind Experiment (SWE) on the Wind
spacecraft \citep[product \texttt{wi\_h1\_swe};][]{Kasper2002}, retrieved using
PySPEDAS \citep{Grimes2022}. This product supplies the parallel and perpendicular
proton thermal speeds, proton density, proton velocity vector, and magnetic field
at a cadence of about $92$~s. We analyze 2004 January~1 to 2008 December~31,
spanning the late declining phase and minimum of solar cycle~23. This phase was
selected because Alfv\'enic slow wind is particularly abundant during declining
and minimum conditions \citep{DAmicisBruno2015,DAmicis2019}, and Wind/SWE
provides a long, consistently sampled interval.

Instrument fill values are replaced with missing data. We retain records with
reduced chi-square $\chi^2_\nu<5$ for the bi-Maxwellian fit, which retains
$93.8$\% of the records, and discard windows in which fewer than half of the
records pass this cut. Intervals associated with interplanetary coronal mass
ejections are removed using the \citet{Richardson2010} ICME catalog, which
identifies 31 ICME periods in 2004--2008. Interplanetary shocks are identified by
a bulk-speed increase exceeding $150\ \mathrm{km\,s^{-1}}$ over three consecutive
records ($\approx4.6$~min), with the flagged records and a $\pm7.7$~min margin
excluded. After transient removal, $29{,}619$ valid one-hour windows remain,
$738$ fewer ($2.4$\%) than without removal.

\subsection{Derived Quantities}
\label{sect:derived}

With the thermal speed defined as $w = (2 k_B T/m_p)^{1/2}$, the proton
anisotropy and parallel beta follow from the SWE fits as
\begin{equation}
\Rp = \frac{T_\perp}{T_\parallel} = \left(\frac{w_\perp}{w_\parallel}\right)^2,
\qquad
\bperp = \frac{2\mu_0 n_p k_B T_\parallel}{B^2}
       = \frac{\mu_0 n_p m_p w_\parallel^2}{B^2}.
\label{eq:plasma}
\end{equation}
For the fluctuation analysis the magnetic field is expressed in Alfv\'en units,
$\mathbf{b} = \mathbf{B}/\sqrt{\mu_0 n_p m_p}$.

\subsection{Stream Classification}
\label{sect:class}

We divide the time series into non-overlapping windows of about one hour (40
records at the $92$~s cadence). For each window we compute the median $\bperp$,
$\Rp$, and bulk speed, together with the normalized cross helicity, normalized
residual energy, magnetic compressive fraction, and normalized field-strength
fluctuation amplitude,
\begin{align}
\sigma_c &=\frac{2\langle\delta\mathbf{v}\mathbin{\cdot}\delta\mathbf{b}\rangle}
 {\langle|\delta\mathbf{v}|^2\rangle+\langle|\delta\mathbf{b}|^2\rangle},
&
\sigma_r &=\frac{\langle|\delta\mathbf{v}|^2\rangle-\langle|\delta\mathbf{b}|^2\rangle}
 {\langle|\delta\mathbf{v}|^2\rangle+\langle|\delta\mathbf{b}|^2\rangle},
\nonumber\\
C_B &=\frac{\langle(\delta|\mathbf{B}|)^2\rangle}
 {\langle|\delta\mathbf{B}|^2\rangle},
&
A_{|B|} &=\frac{\sqrt{\langle(\delta|\mathbf{B}|)^2\rangle}}{B_0}.
\label{eq:fluctuations}
\end{align}
Here $\delta$ denotes the deviation from the window mean and
$B_0=\langle|\mathbf{B}|\rangle$. The quantity $C_B$
runs from 0 for purely transverse fluctuations to 1 for purely compressive
fluctuations and measures a fraction of fluctuation power, whereas $A_{|B|}$
measures the normalized compressive amplitude.

\begin{figure}
\centering
\safeincludegraphics[width=0.9\linewidth]{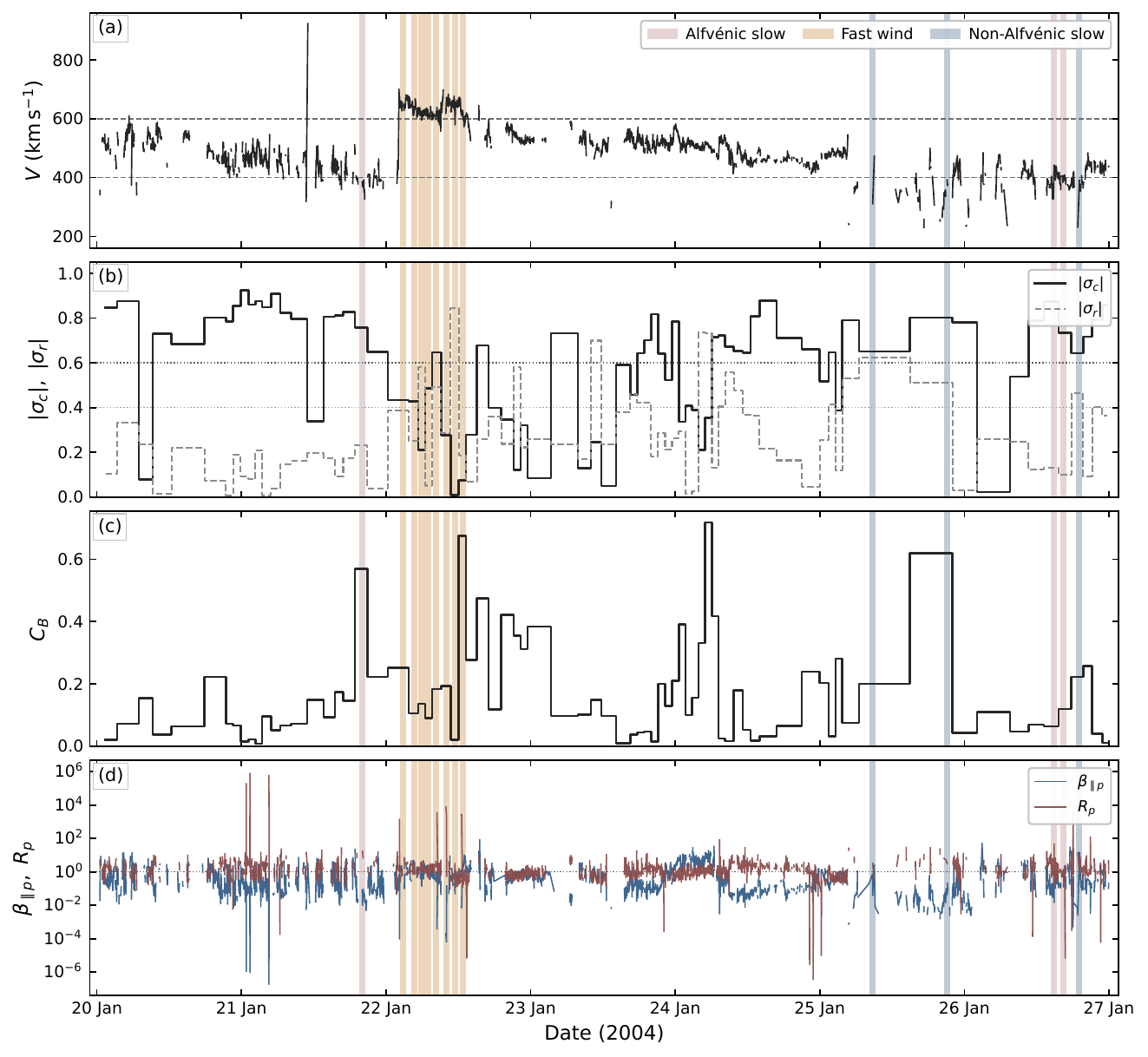}
\caption{Representative seven-day interval (2004 January 20--27) illustrating
the classification procedure. From top to bottom: proton bulk speed $V$
(dashed lines mark the $400$ and $600\ \mathrm{km\,s^{-1}}$ thresholds);
normalized cross helicity $|\sigma_c|$ and residual energy $|\sigma_r|$ per
one-hour window (dotted lines mark the $|\sigma_c|=0.6$ and $|\sigma_r|=0.4$
cuts); magnetic compressive fraction $C_B$; and parallel plasma beta $\bperp$ and
temperature anisotropy $\Rp$ (both per record, log scale; dotted line at
$\Rp=1$). Shaded bands indicate windows classified as fast wind (red),
Alfv\'enic slow wind (dark red), and non-Alfv\'enic slow wind (blue).}
\label{fig:example}
\end{figure}

We classify populations at the one-hour level because $\sigma_c$ and $\sigma_r$
require multiple records. We evaluate instability incidence in two ways: from the
one-hour median state, and from the valid 92-s records assigned their parent
window's population label. The classification criteria are listed in
Table~\ref{tab:classes}, and Figure~\ref{fig:example} illustrates the procedure for
a representative seven-day interval.

\begin{table}[htbp]
\centering
\caption{Stream classification criteria and population sizes for 2004--2008.
Each window is approximately one hour (40 records at $92$~s cadence). The
windows in the intermediate speed range $400$--$600\ \mathrm{km\,s^{-1}}$ are
unclassified and excluded from the population comparison.}
\label{tab:classes}
\begin{tabularx}{\linewidth}{@{}lXr@{}}
\toprule
Population & Criterion & $N$ windows \\
\midrule
Fast & $V > 600\ \mathrm{km\,s^{-1}}$ & $3{,}019$ \\
Alfv\'enic slow & $V < 400\ \mathrm{km\,s^{-1}}$, $|\sigma_c| > 0.6$, $|\sigma_r| < 0.4$ & $2{,}822$ \\
Non-Alfv\'enic slow & $V < 400\ \mathrm{km\,s^{-1}}$, otherwise & $10{,}084$ \\
\bottomrule
\end{tabularx}
\end{table}

\subsection{Instability Margin}
\label{sec:margin}

For the descriptive distributions we overlay the empirical instability thresholds
of \citet{Hellinger2006}. For instability $j$, the fitted threshold anisotropy is
\begin{equation}
R_p^{\mathrm{th},j}(\bperp)
=1+\frac{a_j}{(\bperp-\beta_{0,j})^{b_j}} .
\label{eq:hellinger}
\end{equation}
A boundary is the locus at which the measured anisotropy satisfies
$R_p=R_p^{\mathrm{th},j}$. Equation~(\ref{eq:hellinger}) is an empirical fit and
is evaluated only where it is real and gives $R_p^{\mathrm{th},j}>0$; invalid
branches are masked and carry no physical meaning. Within a valid branch, the
curve marks the mode-specific contour
$\gamma_{j,\max}=10^{-3}\,\Op$. Because some branches are undefined at low beta,
the curves do not enclose a closed stable region. These fits are used for the
descriptive overlays in Figure~\ref{fig:brazil}; every population comparison of
threshold incidence uses the numerically calculated parallel-mode growth rate.

For the parallel-firehose fit, the excluded branch is non-real for
$\bperp<0.59$, singular at $\bperp=0.59$, and non-positive for
$0.59<\bperp\leq0.8306$. The oblique-firehose threshold is non-positive for
$\bperp\leq1.29$.

\begin{table}[htbp]
\centering
\small
\caption{Empirical proton temperature-anisotropy thresholds of
\citet{Hellinger2006} for the mode-specific contour
$\gamma_{j,\max}=10^{-3}\,\Op$. The fits are shown only in the listed domains,
where Eq.~(\ref{eq:hellinger}) is real and yields
$R_p^{\mathrm{th},j}>0$.}
\label{tab:thresholds}
\begin{tabular}{@{}lcccll@{}}
\toprule
Instability & $a_j$ & $b_j$ & $\beta_{0,j}$ & Fit shown for & Anisotropy side \\
\midrule
Proton cyclotron  & $0.43$  & $0.42$ & $-0.0004$ & sampled $\bperp>0$ & $R_p>1$ \\
Mirror            & $0.77$  & $0.76$ & $-0.016$  & sampled $\bperp>0$ & $R_p>1$ \\
Parallel firehose & $-0.47$ & $0.53$ & $0.59$    & $\bperp>0.8306$    & $R_p<1$ \\
Oblique firehose  & $-1.4$  & $1.0$  & $-0.11$   & $\bperp>1.29$      & $R_p<1$ \\
\bottomrule
\end{tabular}
\end{table}

To quantify proximity to instability, we solve the parallel ($k_\perp=0$) bi-Maxwellian Vlasov dispersion relation for a proton population with
charge-neutralizing electrons \citep{Gary1993}. For each plasma state, $\gamma_{\parallel,\max}$ is the largest growth rate over the scanned parallel
wavenumbers for the retained parallel branch: proton cyclotron for $R_p>1$ and parallel firehose for $R_p<1$. The mirror and oblique-firehose modes require 
oblique propagation ($k_\perp\neq0$). Their empirical threshold curves are shown in Figure~\ref{fig:brazil} to indicate their locations in the $(\bperp,R_p)$
plane. Accordingly, $\gamma_{\parallel,\max}$ quantifies proximity to the parallel-mode marginal-stability boundary considered here. The numerical
calculation reproduces the proton-cyclotron and parallel-firehose thresholds of \citet{Hellinger2006} at $10^{-3}\,\Op$ to within 0.22 decades.

The great majority of states lie well below the numerical growth-rate criterion, so the binned median $\gamma_{\parallel,\max}$ is floor-dominated 
and does not discriminate between populations. We therefore use ``marginal stability'' as an operational shorthand for a state at or beyond $\gamma_{\parallel,\max}/\Op=10^{-3}$. 
This definition refers only to the retained parallel modes. At fixed $\bperp$, the threshold incidence measures how often a population reaches this numerical boundary, 
and avoids the ambiguity of a signed distance to the empirical curves near $R_p=1$, where the binding threshold family changes. This fraction, with uncertainty that accounts 
for temporal correlation, is shown for each population in Section~\ref{sect:margin}.

Adjacent one-hour windows are correlated, so uncertainties are estimated with a chronological block bootstrap. We use 16-window blocks, longer than the largest
measured e-folding autocorrelation lag of about 12 windows, and verify that 8-and 32-window blocks give the same ordering. Blocks do not cross data gaps, and
record-level quantities remain grouped within their parent windows. Figures show $68\%$ intervals; contrasts quoted in the text use $95\%$ intervals.

\section{Results}
\label{sect:results}

\subsection{Distribution in the Anisotropy--Beta Plane}
\label{sect:brazil}

\begin{figure}[t]
\centering
\safeincludegraphics[width=0.95\linewidth]{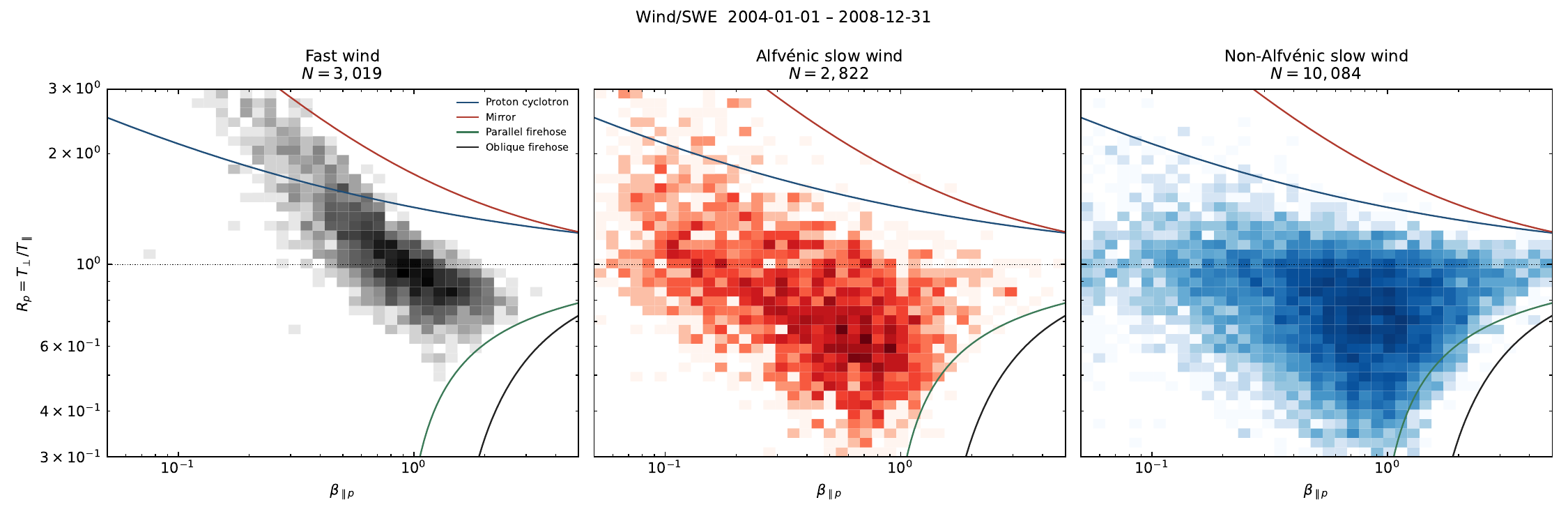}
\caption{Two-dimensional histograms in the plane of parallel proton beta and
temperature anisotropy for the fast wind (left), Alfv\'enic slow wind (center),
and non-Alfv\'enic slow wind (right). Solid curves are the empirical
proton-cyclotron, mirror, parallel-firehose, and oblique-firehose thresholds of
\citet{Hellinger2006}. Only real, positive branches within the validity domains
of Table~\ref{tab:thresholds} are shown; invalid branches are masked. The curves
are descriptive overlays and are not used to calculate the population threshold
incidence.}
\label{fig:brazil}
\end{figure}

Figure~\ref{fig:brazil} shows the three populations in the $(\bperp,\Rp)$
plane with the Hellinger thresholds overlaid. The fast wind and the Alfv\'enic
slow wind both concentrate along the marginal stability ridge, while the
non-Alfv\'enic slow wind occupies a broader, more isotropic region. Population
sizes are listed in Table~\ref{tab:classes}.

Table~\ref{tab:summary} gives a compact physical comparison of the three
populations. Each entry is the median across windows with the interquartile
range in brackets.

\begin{table}[htbp]
\centering
\scriptsize
\caption{Population summary statistics. Here
$\rho_i=w_\perp/\Omega_p$ is the perpendicular proton thermal gyroradius and
$\omega_{pp}$ is the proton plasma angular frequency. The field $B_0$ is the
mean field magnitude in each window, and $M_A=V/v_A$.}
\label{tab:summary}
\begin{tabularx}{\linewidth}{@{}l>{\centering\arraybackslash}X>{\centering\arraybackslash}X>{\centering\arraybackslash}X@{}}
\toprule
Quantity & Fast wind & Alfv\'enic slow & Non-Alfv\'enic slow \\
\midrule
$V$ (km s$^{-1}$) & $643.9$ [$621.8$, $673.6$] & $354.8$ [$329.4$, $377.5$] & $345.8$ [$320.2$, $370.0$] \\
$n_p$ (cm$^{-3}$) & $2.424$ [$2.001$, $3.108$] & $5.240$ [$3.788$, $7.687$] & $7.098$ [$4.831$, $10.452$] \\
$B_0$ (nT) & $4.576$ [$3.889$, $5.665$] & $4.787$ [$3.613$, $6.088$] & $3.925$ [$2.925$, $5.387$] \\
$v_A$ (km s$^{-1}$) & $64.42$ [$56.10$, $76.23$] & $46.35$ [$35.59$, $58.73$] & $33.60$ [$25.04$, $44.46$] \\
$M_A$ & $10.06$ [$8.53$, $11.57$] & $7.71$ [$6.03$, $9.83$] & $10.32$ [$7.91$, $13.44$] \\
$\bperp$ & $0.770$ [$0.481$, $1.130$] & $0.432$ [$0.196$, $0.728$] & $0.702$ [$0.395$, $1.092$] \\
$R_p$ & $1.020$ [$0.868$, $1.318$] & $0.782$ [$0.599$, $1.027$] & $0.774$ [$0.612$, $0.945$] \\
$\rho_i$ (km) & $128.2$ [$110.9$, $147.2$] & $53.6$ [$41.9$, $68.0$] & $58.3$ [$44.6$, $75.3$] \\
$\omega_{pp}$ (rad s$^{-1}$) & $2050$ [$1862$, $2321$] & $3013$ [$2562$, $3650$] & $3508$ [$2894$, $4256$] \\
$C_B$ & $0.0203$ [$0.0110$, $0.0374$] & $0.0219$ [$0.0093$, $0.0602$] & $0.0436$ [$0.0186$, $0.1021$] \\
$A_{|B|}$ & $0.0745$ [$0.0501$, $0.1082$] & $0.0418$ [$0.0257$, $0.0718$] & $0.0738$ [$0.0422$, $0.1283$] \\
$N$ windows & $3{,}019$ & $2{,}822$ & $10{,}084$ \\
\bottomrule
\end{tabularx}
\end{table}

\subsection{Compressive Fraction and Amplitude}
\label{sect:compress}

\begin{figure}
\centering
\safeincludegraphics[width=0.9\linewidth]{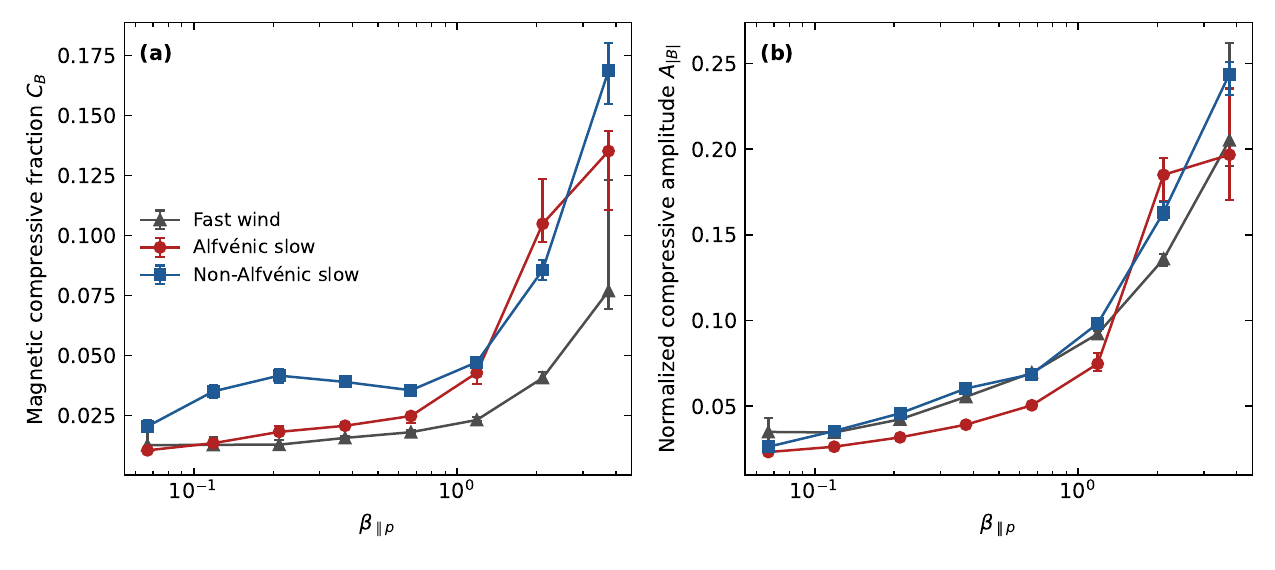}
\caption{Binned median (a) magnetic compressive fraction $C_B$ and (b)
normalized field-strength fluctuation amplitude $A_{|B|}$ as functions of
$\bperp$ for the three populations. Error bars are chronological block-bootstrap
$68\%$ intervals.}
\label{fig:compress}
\end{figure}

Figure~\ref{fig:compress} separates magnetic compressive fraction $C_B$ from
normalized compressive amplitude $A_{|B|}$. The global median $C_B$ is $0.022$
in the Alfv\'enic slow wind and $0.044$ in the non-Alfv\'enic slow wind; these
values describe a fraction of fluctuation power, not its amplitude. The
corresponding median amplitudes are $0.0418$ and $0.0738$. At matched
$\bperp<0.4$, the median amplitude is lower in the Alfv\'enic slow wind in all
four bins, with the $95\%$ interval excluding zero in three. This supports, but
does not establish, the fluctuating-anisotropy interpretation.

\subsection{Proximity to Marginal Stability}
\label{sect:margin}

\begin{figure}
\centering
\safeincludegraphics[width=0.9\linewidth]{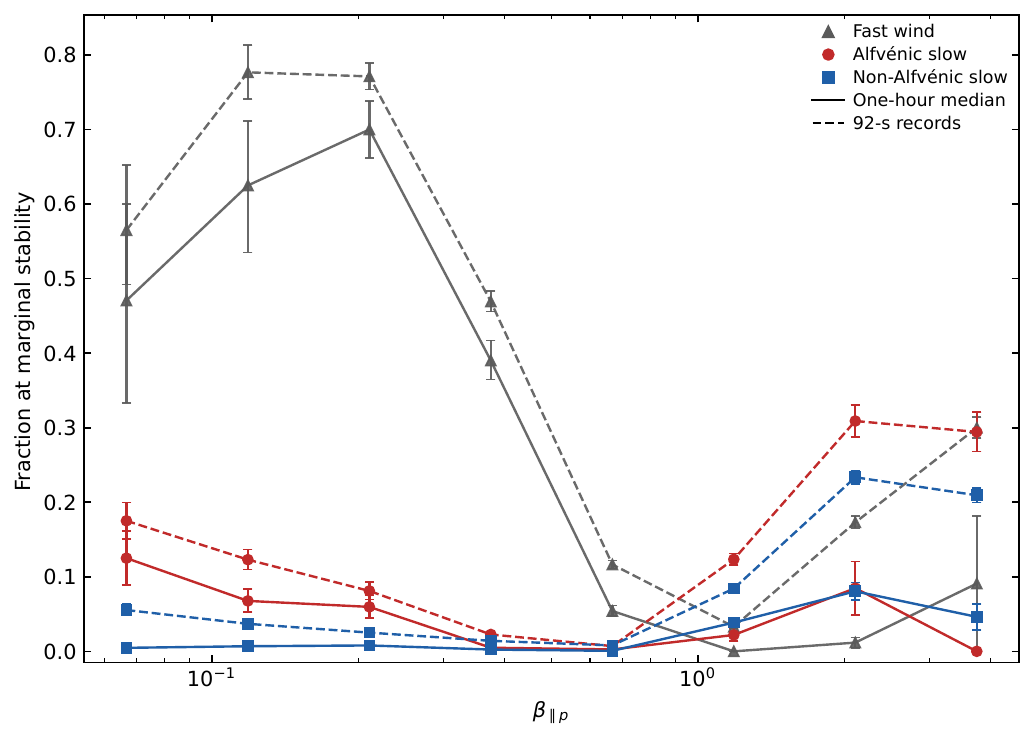}
\caption{Fraction at the operational parallel-mode marginal-stability criterion,
$\gamma_{\parallel,\max}/\Op\geq10^{-3}$, as a function of $\bperp$. Solid
curves use the one-hour median state; dashed curves evaluate every valid 92-s
record while retaining the parent window's population label. Error bars are
chronological block-bootstrap $68\%$ intervals.}
\label{fig:margin}
\end{figure}

Figure~\ref{fig:margin} compares the threshold incidence obtained from each
window's median state with that obtained from every valid 92-s record assigned
its parent window's population label. The record-level analysis identifies
within-window crossings missed by the hourly medians in both populations. In the
three lowest beta bins, however, threshold incidence remains higher in the
Alfv\'enic slow wind at both levels, and the block-bootstrap $95\%$ intervals for
the Alfv\'enic-minus-non-Alfv\'enic differences remain above zero. All low-beta
window-level crossings lie on the $R_p>1$ proton-cyclotron side. Above
$\bperp\simeq1$, the Alfv\'enic sample is too sparse to extend this comparison.

\subsection{Continuity with Alfv\'enicity and robustness to the cross-helicity cut}
\label{sect:continuity}

\begin{figure}
\centering
\safeincludegraphics[width=0.7\linewidth]{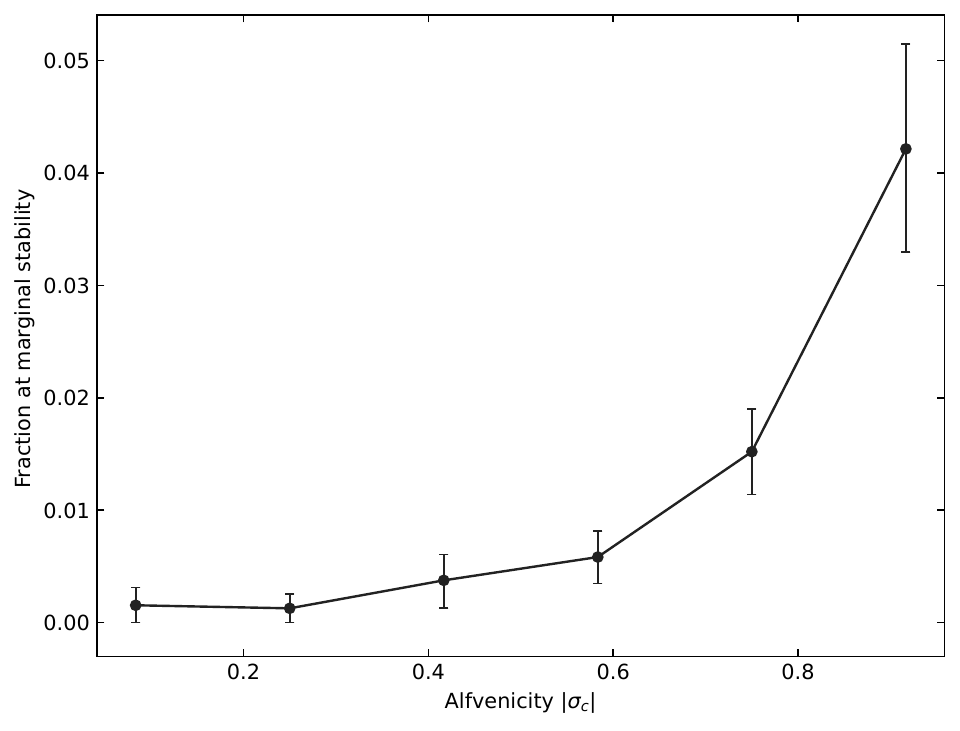}
\caption{Fraction of windows at marginal stability ($\gamma_{\parallel,\max}/\Op \geq
10^{-3}$) as a function of the Alfv\'enicity $|\sigma_c|$, for the pooled slow
wind ($V < 400\ \mathrm{km\,s^{-1}}$) restricted to the parallel-beta band
$0.1 \leq \bperp < 0.6$ to control for the beta dependence. Error bars are
chronological block-bootstrap $68\%$ intervals. The overall rise shows that the
result is not confined to the cross-helicity threshold used in
Figure~\ref{fig:margin}.}
\label{fig:continuity}
\end{figure}

\begin{figure}
\centering
\safeincludegraphics[width=\linewidth]{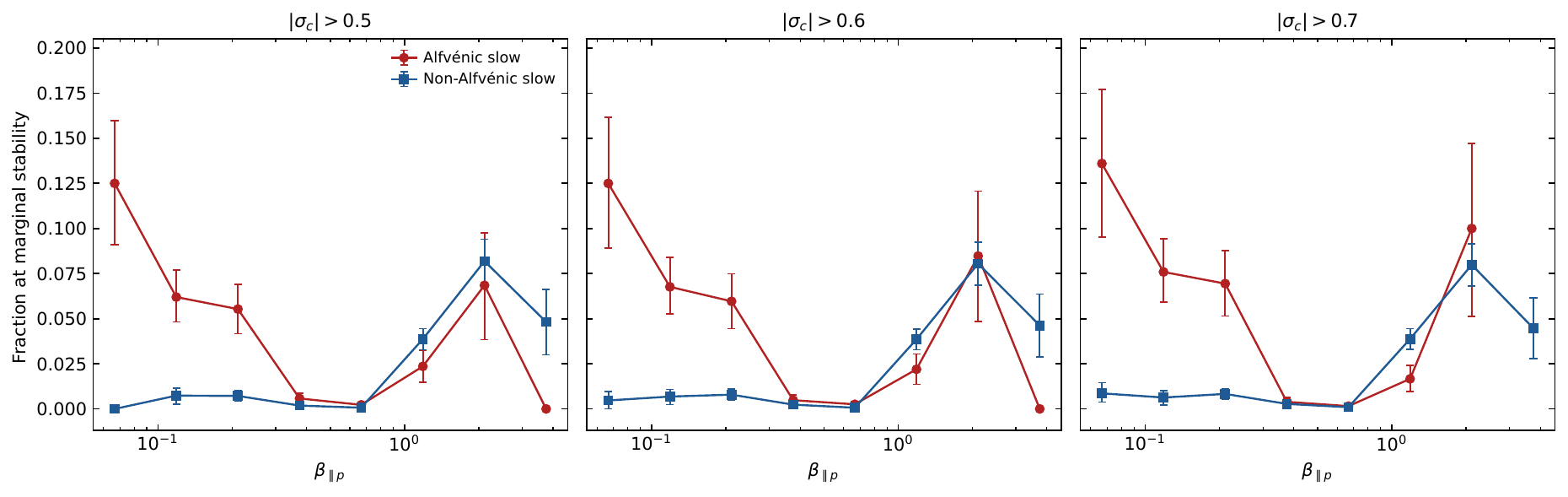}
\caption{Robustness of the matched-$\bperp$ result to the cross-helicity cut. Each
panel repeats the Figure~\ref{fig:margin} comparison of the Alfv\'enic and
non-Alfv\'enic slow wind with the cross-helicity threshold set to $|\sigma_c| >
0.5$, $0.6$, and $0.7$ (the residual-energy cut $|\sigma_r| < 0.4$ is kept
throughout). The ordering---the Alfv\'enic slow wind closer to marginal
stability in the three lowest beta bins---is unchanged. Error bars are
chronological block-bootstrap $68\%$ intervals.}
\label{fig:robust}
\end{figure}

The contrast between the two slow populations is robust to the tested
cross-helicity cuts. Figure~\ref{fig:continuity} pools all slow-wind windows in
the controlled band $0.1 \leq \bperp < 0.6$ and bins them by Alfv\'enicity
$|\sigma_c|$: the fraction at marginal stability rises overall from
$\lesssim 0.005$ for the least Alfv\'enic windows to $\approx 0.04$ for the most
Alfv\'enic. This trend shows that the result is not confined to a single
cross-helicity threshold. Consistently, the matched-$\bperp$ ordering of Figure~\ref{fig:margin} is
preserved when the cross-helicity cut is moved between $|\sigma_c| > 0.5$ and
$|\sigma_c| > 0.7$ (Figure~\ref{fig:robust}): in every case the Alfv\'enic slow
wind has greater threshold incidence in the three lowest beta bins.

\subsection{Spectral compressive fraction in representative intervals}
\label{sect:psd}

\begin{figure}
\centering
\safeincludegraphics[width=\linewidth]{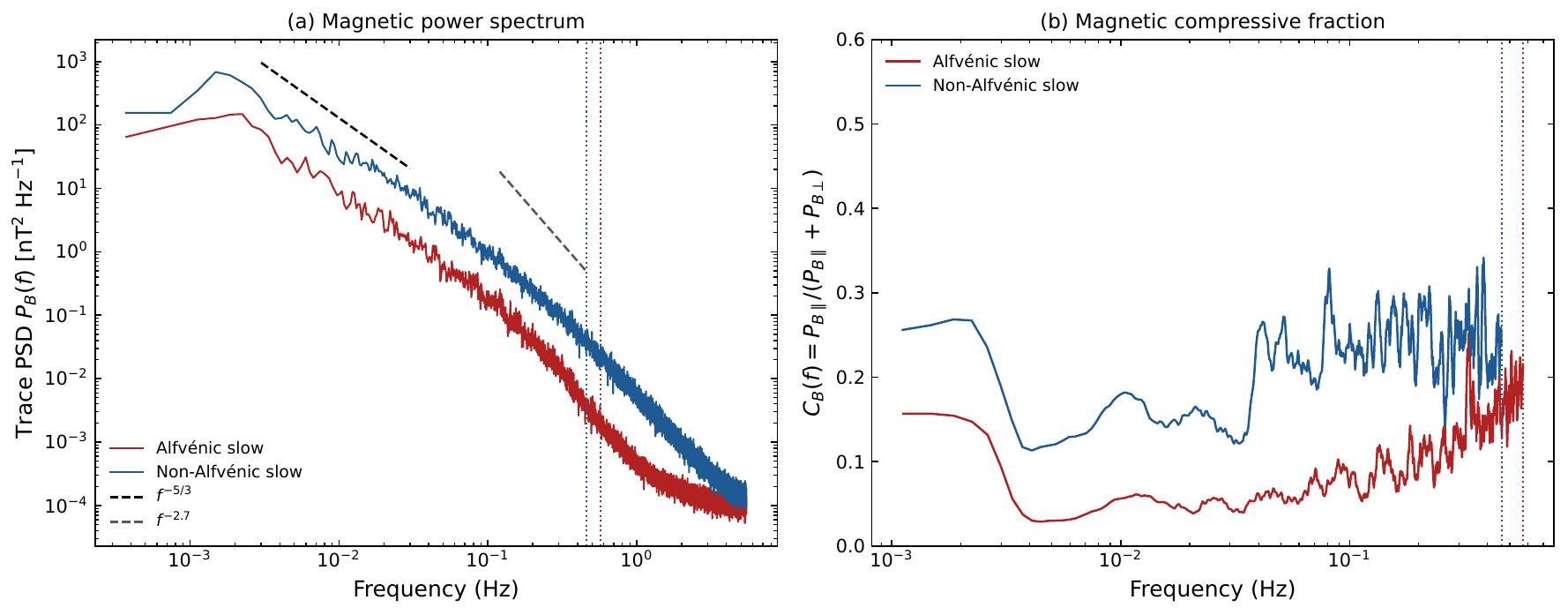}
\caption{Qualitative spectral comparison of representative Alfv\'enic (2004
March~7) and non-Alfv\'enic (2004 February~24) slow-wind intervals from Wind/MFI
high-resolution data. (a) Trace power spectrum $P_B(f)$, with $f^{-5/3}$
(inertial) and $f^{-2.7}$ (kinetic) reference slopes and the ion-inertial
frequency $f_{di}$ marked by dotted vertical lines. (b) Frequency-resolved
magnetic compressive fraction $C_B(f)$. The examples illustrate a difference in
compressive fraction; because $C_B(f)$ is a ratio, they do not establish a
difference in normalized compressive amplitude.}
\label{fig:psd}
\end{figure}

Figure~\ref{fig:psd} provides a qualitative spectral comparison of two
representative intervals. It shows the trace magnetic power spectrum and the
frequency-resolved magnetic compressive fraction
$C_B(f)=P_{B\parallel}/(P_{B\parallel}+P_{B\perp})$ for a six-hour Alfv\'enic
interval (2004 March~7, $00$--$06$~UT) and a non-Alfv\'enic interval (2004
February~24, $06$--$12$~UT), computed from Wind/MFI high-resolution
magnetic-field data ($\approx11$~vectors\,s$^{-1}$) in a field-aligned coordinate
system. Both intervals show an inertial-range slope close to $f^{-5/3}$ that
steepens toward the kinetic range near the ion-inertial frequency $f_{di}$. Across
much of the plotted inertial range, $C_B(f)$ is lower in the Alfv\'enic example
than in the non-Alfv\'enic example. Because $C_B(f)$ is a ratio and the two
intervals are illustrative rather than a population sample,
Figure~\ref{fig:psd} is not used as evidence for an ordering of normalized
compressive driving amplitude.

\section{Discussion}
\label{sect:discussion}

At low beta, the Alfv\'enic slow wind has lower normalized field-strength
fluctuation amplitude and greater parallel-mode threshold incidence than the
non-Alfv\'enic slow wind. This association is consistent with the
fluctuating-anisotropy picture of \citet{Verscharen2016}, in which weaker
compressive driving permits the mean plasma state to remain closer to a
stability boundary. The observational comparison does not establish causation,
so we present this mechanism only as an interpretation.

The anisotropy of the marginal windows clarifies which retained parallel branch
is relevant. Every marginal window in the three lowest beta bins lies at
$R_p>1$ on the proton-cyclotron side, and no positive $R_p<1$
parallel-firehose contour exists there at $10^{-3}\,\Op$. The low-beta result
therefore does not indicate firehose-side regulation.

Our measurement is made at 1~au, and its radial evolution remains to be tested.
The proton anisotropy is already constrained by instability thresholds in the
inner heliosphere and remains constrained as the wind expands
\citep{Matteini2007}, while the turbulent compressive fraction generally evolves
as Alfv\'enic correlations decay. Measurements of Alfv\'enic slow wind with
Helios, Parker Solar Probe, and Solar Orbiter
\citep{Stansby2020,Perrone2020} provide a direct test of the association at
smaller heliocentric distances. Coulomb collisions also isotropize protons, so a
collisional contribution to the observed ordering cannot be excluded.

The 2004--2008 selection limits the present inference to the declining and minimum phases of solar cycle~23. 
Statistical studies show that the Alfv\'enic content and occurrence of Alfv\'enic slow wind vary with solar-cycle phase, 
while ICME occurrence also changes substantially over the cycle \citep{DAmicis2021,Richardson2010}. 
A solar-maximum sample may therefore differ in both population composition and transient contamination. 
The present data set does not determine whether the matched-$\bperp$ association reported here is 
present with the same strength, or at all, during solar maximum. We therefore do not predict whether 
the association should strengthen or weaken. Establishing this will require applying the same population 
classification and threshold-incidence analysis to an independent solar-maximum interval.

Pressure-anisotropy instabilities and mixtures of Alfv\'enic and compressive
fluctuations also occur in other weakly collisional plasmas, such as the hot
intracluster medium and collisionless accretion flows. Whether a similar
association occurs in those systems remains to be tested.

Several limitations should be noted. The numerical thresholds are calculated for
a proton--electron plasma. Alpha particles can shift the firehose and mirror
limits \citep{Chen2016}, but their effect on the population ordering is not
quantified by the present solver. The bi-Maxwellian description also neglects
field-aligned proton beams, which are common in faster streams and can modify
both the inferred anisotropy and the instability thresholds
\citep{Marsch2006}. The analysis is performed at a single heliocentric distance
and depends on the chosen speed, cross-helicity, and residual-energy cuts.
Figure~\ref{fig:robust} tests nearby cross-helicity cuts while keeping
$|\sigma_r|<0.4$ fixed and preserves the reported ordering over that tested
range; sensitivity to the speed and residual-energy cuts has not been evaluated.

\section{Summary and Conclusions}
\label{sect:summary}

The analysis establishes an empirical association between Alfv\'enicity and how
frequently the slow solar wind approaches the parallel-mode marginal-stability
criterion at matched $\bperp$ during 2004--2008. Magnetic compressive fraction
and normalized compressive amplitude are distinct observables, and the latter is
the quantity relevant to the fluctuating-anisotropy interpretation. This
interpretation is presented as a possible explanation rather than a causal
conclusion. Extending the analysis across heliocentric distance and solar-cycle
phase will test the generality of the association.

\FloatBarrier

\section*{Acknowledgements}
We thank the Wind/SWE instrument team and NASA's Space Physics Data Facility
for providing the data used in this study.

\section*{Declaration of competing interest}
The authors declare that they have no known competing financial interests or
personal relationships that could have appeared to influence the work reported
in this paper.

\section*{Data availability}
The Wind/SWE bi-Maxwellian proton data (product \texttt{wi\_h1\_swe}) are
publicly available from NASA's Space Physics Data Facility (CDAWeb,
\url{https://cdaweb.gsfc.nasa.gov}). The interplanetary coronal mass ejection
catalog of \citet{Richardson2010} is publicly available. The analysis code that
produced the figures is available from the corresponding author on reasonable
request.

\FloatBarrier

\bibliographystyle{elsarticle-harv}
\bibliography{references}

\end{document}